\documentclass{article}
\usepackage{cite} 

\usepackage[english]{babel}

\usepackage[letterpaper,top=2cm,bottom=2cm,left=3cm,right=3cm,marginparwidth=1.75cm]{geometry}

\usepackage{amsmath}
\usepackage{graphicx}
\usepackage[colorlinks=true, allcolors=blue]{hyperref}

\title{\textbf{Role of Databases in GenAI Applications}}
\author{Santosh Bhupathi Sr. Solutions Architect bhupathi.santosh@gmail.com}
\date{}
\begin{document}

\maketitle

\begin{abstract}
Generative AI (GenAI) is transforming industries by enabling intelligent content generation, automation, and decision making. However, the effectiveness of GenAI applications is significantly dependent on efficient data storage, retrieval, and contextual augmentation. This paper explores the critical role of databases in GenAI workflows, emphasizing the importance of choosing the right database architecture to optimize performance, accuracy, and scalability. It categorizes database roles into conversational context (key-value/document databases), situational context (relational databases/data lakehouses), and semantic context (vector databases), each serving a distinct function in enriching AI-generated responses. In addition, the paper highlights real-time query processing, vector search for semantic retrieval, and the impact of database selection on model efficiency and scalability. Using a multi-database approach, GenAI applications can achieve more context-sensitive, personalized, and high-performance AI-driven solutions.
\end{abstract}

\section*{\textbf{Introduction (Generative AI \& Large Language Models)}}

Generative AI (GenAI) represents a transformative leap in artificial intelligence, using advanced models such as Transformers, GPT-4, and Gemini to generate human-like content in multiple modalities [1],[2]. Unlike traditional AI models that focus on classification or predictive tasks using predefined patterns, GenAI utilizes deep learning architectures like Transformer-based Large Language Models (LLMs)[2] to create text, images, code, and audio.
The most prominent GenAI models include GPT-4 for advanced text generation[1] and Google Gemini for multimodal AI applications[2]. These models leverage massive data sets and training methodologies such as Reinforcement Learning with Human Feedback (RLHF)[3] and retrieval-augmented generation (RAG)[4] to improve their contextual understanding and adaptability.
These AI models, trained on large-scale data, can understand context, generate creative outputs, automate workflows, and drive innovation across industries. GenAI is transforming fields such as healthcare (AI-assisted diagnosis and drug discovery[5]), finance (automated risk analysis and fraud detection[6]), customer support (intelligent virtual assistants[7]), and software development (AI-driven code generation[8]). The emergence of multimodal AI which enables models to process and generate text, images, and audio simultaneously is further unlocking new possibilities in automation, personalization, and decision-making.

\section*{Value Proposition of Generative AI Applications}
Generative AI applications deliver substantial business value by enhancing efficiency, creativity, and decision-making. Below are some key value propositions:

\begin{enumerate}
    \item \textbf{Enhanced Productivity \& Automation}

\begin{itemize}
    \item Automates repetitive tasks like document generation, summarization, and code completion, reducing manual effort and improving efficiency.
    \item Enables self-service customer support with AI-powered chatbots that provide contextual and natural interactions.
\end{itemize}

\item \textbf{Personalized User Experience}

\begin{itemize}
    \item Powers hyper-personalization by generating tailored content, recommendations, and responses based on user preferences and context.
    \item Enhances marketing efforts with AI-generated ad copies, email campaigns, and personalized product recommendations.
\end{itemize}

\item \textbf{Intelligent Decision-Making}

\begin{itemize}
    \item Helps in real-time data analysis and insights extraction for better decision-making in finance, healthcare, and operations.
    \item Improves fraud detection, risk analysis, and compliance monitoring by analyzing vast amounts of structured and unstructured data.
\end{itemize}

\item \textbf{Creativity \& Content Generation}

\begin{itemize}
    \item Generates high-quality content, including articles, product descriptions, scripts, and creative writing for media and marketing.
    \item Assists in design, music, and video generation, reducing creative bottlenecks and accelerating production cycles.
\end{itemize}

\item \textbf{Cost Optimization \& Scalability}

\begin{itemize}
    \item Reduces costs associated with manual content creation, customer support, and software development by automating key processes.
    \item Scales seamlessly across global operations, enabling faster market expansion and localized content generation.
\end{itemize}

\item \textbf{Democratization of AI \& Innovation}

\begin{itemize}
    \item Enables non-technical users to leverage AI-driven tools for content creation, analysis, and automation.
    \item Empowers developers, researchers, and enterprises to build new AI-driven applications without extensive AI expertise.
\end{itemize}
\end{enumerate}

\section*{From Bits to Brains: Why the Right Database Matters in GenAI}
In many GenAI applications, there’s a common misconception that only a vector database is needed, particularly for semantic search and embedding management[9],[10]. However, different parts of the application often require specialized storage solutions. Vector databases undoubtedly play a pivotal role in semantic context, but relying on them exclusively can lead to performance bottlenecks, data inconsistencies, and an incomplete view of user interactions[11]. By matching each data type to the most appropriate database technology, GenAI solutions can deliver faster, more accurate, and contextually rich responses, ensuring the true potential of AI-driven applications is fully realized[12].

In GenAI applications, real-time data processing demands databases that can handle rapid queries and large-scale ingestion without bottlenecks [13]. Traditional database systems often struggle with the high throughput required to train and deploy AI models, making scalability and low latency critical. Continuous model updates also mean that databases must support efficient retrieval mechanisms to ensure model integrity. Ultimately, selecting the wrong database can lead to performance degradation, compromised accuracy, and diminished overall effectiveness of GenAI solutions.

\section*{Role of Databases in Generative AI Applications}
Databases play a critical role in Generative AI (GenAI) applications, ensuring efficient data storage, retrieval, and contextual augmentation for AI-driven responses[14],[15]. When a user interacts with a GenAI-powered system, the application relies on multiple databases to fetch historical context, enrich responses with enterprise knowledge, and enhance overall accuracy.

In the user's critical path, databases contribute to key stages of the workflow, enabling AI models to make informed decisions by leveraging structured and unstructured data. When an end-user interacts with a Generative AI application, they typically post a question or prompt to the system. At first glance, this may seem like a simple request-response mechanism, but behind the scenes, a complex orchestration of databases and AI models takes place to deliver accurate, context-aware, and relevant responses[15].

In this blog, we’ll explore how databases play a crucial role in enhancing Generative AI applications by providing different types of contextual data that shape the final response. Below is an overview of how databases are used at different steps of a GenAI interaction.

\section*{How Databases Support the User’s Critical Path in GenAI Applications}

\begin{figure}[h!]
    \centering
    \includegraphics[width=1.10\linewidth]{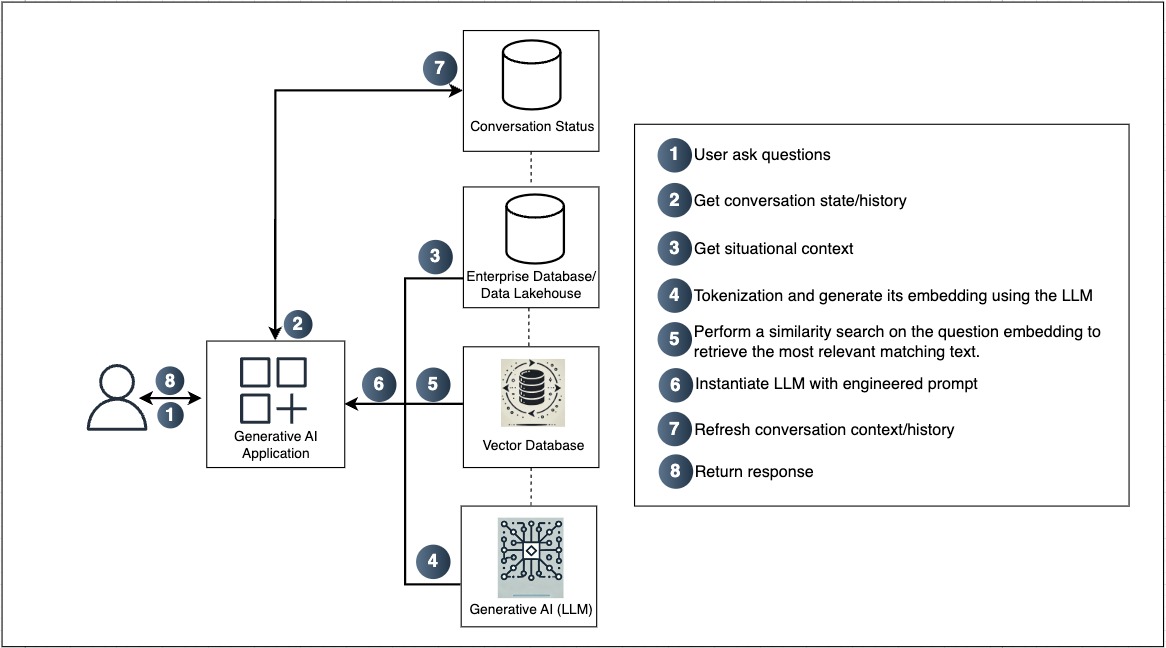}
    \caption{Here is a high-level workflow of a GenAI Application.}
    \label{fig:Here is a high-level workflow of a GenAI Application.}
\end{figure}

\begin{itemize}
\item\textbf{Step 1: User Asks Questions}
\end{itemize}

The user initiates an interaction by submitting a query, request, or instruction to the GenAI application. This can range from simple questions to more complex tasks requiring in-depth analysis or content generation. The system captures this input and sets the stage for subsequent steps, ensuring it has a clear starting point for generating a relevant, context-aware response.

\textbf{Database Role:}
\begin{itemize}
    \item At this point, no direct database interaction is strictly required however, the question itself will later be stored or logged for future reference and to maintain conversation continuity. Typically, the application logs the user’s query in a database for potential auditing, analytics, or future reference. If user or session data is required at this point such as user authentication or preferences then the system may query a relational or key-value store to validate or personalize the request before moving to the next steps.
\end{itemize}

\begin{itemize}
\item \textbf{Step 2: Understanding User Interaction: Query Processing \& Contextual Data Retrieval}
\end{itemize}
    
The user submits a query to the Generative AI application, as soon as the user asks a question, the application loads a relevant prompt template. This template engineering process enhances the original question by adding additional context, ensuring the Large Language Model (LLM) produces a more accurate and relevant output.

\textbf{Conversational Context (Maintaining Chat History)} refers to the process of preserving and managing all previous interactions between the user and the system. This includes user queries, the system’s responses, and any relevant metadata or context. By storing and referencing this historical data in an enterprise database or data Lakehouse application can provide more coherent and personalized responses in subsequent interactions. Essentially, it enables the AI to “remember” what was previously discussed, thereby improving continuity and user experience in a conversational setting.

\textbf{Database Role:}
\begin{itemize}
    \item Since LLMs do not retain memory, the system stores previous user interactions in a database.
    \item This ensures continuity in conversations, making the AI responses more coherent and contextually relevant over multiple interactions.
    \item Enterprise Database / Data Lakehouse: Stores long-term conversation data or user session logs. When the application needs the historical context such as the user’s previous questions, system instructions, or any relevant session metadata it queries this database.
    \item Why It Matters: Persisting conversation data ensures the application can reference past interactions, enabling more coherent, context-aware responses.
\end{itemize}

\begin{itemize}
    \item \textbf{Step 3: Context Enrichment \& Knowledge Retrieval}
\end{itemize}

Large Language Models (LLMs) are inherently stateless, treating each query as if it were their first. They lack any built-in mechanism to recall previous interactions or shared information, making an external data store indispensable for preserving context. Databases fill this need by retaining conversation history, user profiles, and supplementary knowledge ultimately enriching the AI’s ability to generate coherent, context-aware responses.

\textbf{Situational Context (User Profile \& Operational Data)
}This involves gathering any additional information the system may need to provide a well-informed response. Examples include user profile details (such as preferences or history), operational data (like current system status or relevant business metrics), and any domain-specific knowledge. By querying an enterprise database or data Lakehouse at this stage, the application retrieves the contextual data needed to tailor responses accurately and ensure relevance.

\textbf{Database Role:}
\begin{itemize}
    \item The application also needs user-specific and real-time data to generate a personalized response.
    \item If the system requires sub-millisecond latency, A caching service is used for fast retrieval of frequently accessed data.
    \item The system queries a conversation history database to retrieve past interactions and user preferences.
    \item If the AI application supports personalization, relational databases may fetch user-specific data to tailor responses.
\end{itemize}

\begin{itemize}
    \item \textbf{Step 4 \& 5: LLM-Based Embeddings and Vector Search: From Tokenization to Semantic Retrieval}
\end{itemize}
    
The application tokenizes the original question, converting the text into a numerical form known as embeddings. These embeddings, generated by the large language model (LLM), capture the semantic meaning of the user's query. The embeddings serve as a representation of the question in a way that helps the system understand the underlying context, enabling more accurate and contextually relevant responses. The process of embedding generation using the LLM facilitates the transformation of raw text into meaningful, machine-readable representations essential for downstream tasks.

\textbf{Semantic Context: Tokenization and Embedding Generation with the LLM)}
The application begins by converting the user’s query into a sequence of tokens effectively translating text into a numerical form. This process, known as embedding generation, leverages a large language model (LLM) to produce high-dimensional vector representations that capture semantic meaning. By creating these embeddings, the LLM provides a nuanced context for the user’s question, laying the groundwork for more accurate and context-aware responses.

\textbf{Similarity search (Vectorized Knowledge Retrieval)}
Perform a Similarity Search on the Question Embedding in the GenAI workflow. Here, the system leverages a vector database to retrieve semantically similar documents or data points by comparing the numerical embeddings generated in the previous step. This allows the application to provide contextually rich and relevant information essentially capturing the “meaning” of the user’s query rather than relying solely on keyword matches.
At this stage, all three types of contexts (conversational, situational, and semantic) are synthesized into an engineered prompt to provide the LLM with the best possible input. This enables semantic search, allowing the model to find relevant information even when the exact words differ.

\textbf{Database Role:}
\begin{itemize}
    \item To enhance understanding, the AI converts the user’s query into embeddings (mathematical representations of text). Need a specialized database called vector database.
    \item These embeddings are then searched against a Vector Database to retrieve similar text or knowledge snippets.
    \item Knowledge Graphs and vector databases assist in semantic search, allowing the model to find relevant facts and documents before generating a response.
    \item Vector databases perform similarity searches on pre-indexed text, images, or other embeddings to find relevant context for the AI model.
    \item This step is essential for RAG (Retrieval-Augmented Generation), where AI models ground their responses with real-time or proprietary knowledge instead of relying solely on pre-trained data.
\end{itemize}

\textbf{Understanding Vector Databases: The Engine Behind Semantic Search}
A vector database is a specialized data store designed to handle high-dimensional numerical vectors, often called embeddings. A vector database indexes and stores vector embeddings for fast retrieval and similarity search. In GenAI applications, these embeddings represent the semantic meaning of text, images, or other data types [16]. By storing and indexing these embeddings, vector databases enable efficient similarity searches based on conceptual proximity rather than exact string matches.

\textbf{Why Use a Vector Database for Semantic or Similarity Search?}
\begin{enumerate}
    \item \textbf{High-Dimensional Data:} GenAI models frequently produce embeddings that can have hundreds or thousands of dimensions. Vector databases provide specialized indexes (e.g., IVF, HNSW) optimized for performing approximate nearest neighbor (ANN) searches at scale[18].
    \item \textbf{Semantic Matching:} Instead of matching exact keywords, vector databases identify conceptually related results even if the words differ. This is essential in scenarios where synonyms or paraphrased text must be recognized as relevant.
    \item \textbf{Low Latency at Scale:} Vector databases are built to handle large volumes of embeddings while still delivering fast query times. Traditional databases often struggle with performance when tasked with similarity searches in high-dimensional space.
    \item \textbf{Integration with GenAI Pipelines:} By storing model-generated embeddings, vector databases streamline the retrieval step in many GenAI workflows, such as retrieving semantically related documents for question answering or recommendation systems.
\end{enumerate}

\textbf{Why Not Use a Typical Relational or NoSQL Database?}
\begin{enumerate}
    \item \textbf{Lack of Specialized Indexing:} Relational and many NoSQL databases are optimized for structured data lookups or key-based queries, not high-dimensional similarity searches. They do not typically support the specialized indexing structures needed for efficient ANN operations.
    \item \textbf{Performance Bottlenecks:} Attempting to store and query large numbers of embeddings in a standard database can lead to high latency and resource usage, as these systems are not designed for vector-based lookups[18].
    \item \textbf{Limited Query Capabilities:} While you can force a relational or NoSQL database to store embeddings, you’d still need an external library or application logic to perform similarity searches. This approach is often cumbersome, less efficient, and difficult to scale.
    \item \textbf{Scaling Complexities:} Vector databases are built with distributed architectures and optimized data structures specifically for scaling similarity search workloads. Adapting a relational or general-purpose NoSQL store for these tasks can become unwieldy or prohibitively expensive.
\end{enumerate}

Let’s explore this vector database with an example use-case of an e-commerce store selling running shoes, where we’ll see how semantic search helps users find the perfect pair within their budget.

\textbf{The Problem: Keyword Search vs. Semantic Search}

Many e-commerce platforms rely on keyword-based search engines. While these systems can handle exact matches (e.g., "red running shoes"), they often miss nuanced queries like "comfortable sneakers under \$100." This gap leads to missed opportunities and a poor user experience.
Why It Matters
\begin{itemize}
    \item Enhanced User Satisfaction: Users find products that truly match their intent.
    \item Improved Conversions: More relevant results lead to higher purchase rates.
\end{itemize}

\textbf{Sample Inventory Data}

Consider a running shoe inventory. For simplicity, each product is represented by a 2D vector embedding (real embeddings often have hundreds of dimensions). 

Here’s our sample inventory in tabular form:
\begin{figure}[h!]
    \centering
    \includegraphics[width=1.10\linewidth]{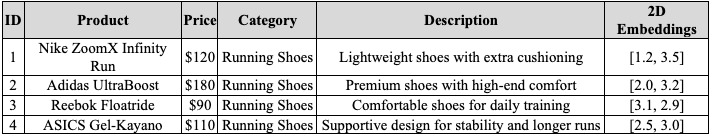}
    \caption{Table of Running Shoes with Descriptions and 2D Embeddings}
    \label{fig:able of Running Shoes with Descriptions and 2D Embeddings}
\end{figure}

    \textbf{User Query}
A user visits the e-commerce site and types: “I need comfortable running shoes under \$100.”
Then A text embedding model (like Sentence-BERT) converts this query into a 2D embedding:

\begin{figure}[h]
    \centering
    \includegraphics[width=1.10\linewidth]{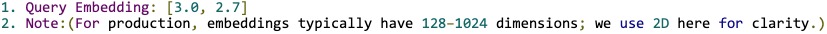}
    \caption{2D Embeddings}
    \label{fig:2D Embeddings}
\end{figure}

\textbf{Vector Database Search}

Instead of relying on keyword matching, we store product embeddings in a vector database (such as Milvus, Pinecone, or Vespa). When the user submits the query:

\begin{itemize}
    \item The application generates the embedding of the query [3.0, 2.7].
    \item The vector database performs a nearest-neighbor search to find the most semantically similar products.
    \item Metadata filters (e.g., “under \$100”) are applied to ensure that only budget-friendly items are returned.
\end{itemize}

A simplified flow diagram illustrates the process.
\\
\begin{figure}[h!]
    \centering
    \includegraphics[width=1.0\linewidth]{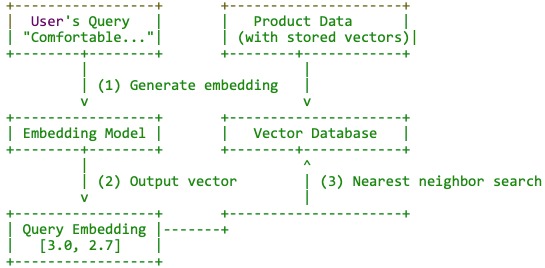}
    \caption{Vector search workflow}
    \label{fig:Vector search workflow}
\end{figure}
\\
\\

\textbf{Distance Calculations (Euclidean)[19]}
\\
To determine the similarity between the query and each product, we use Euclidean distance, which is the straight-line distance between two points:

\centering distance(A,B)= $\sqrt{(xA-xB)^2+(yA-yB)^2}$

Calculated distances from the query embedding [3.0, 2.7] to each product are as follows:

\begin{enumerate}
        \item Nike ZoomX Infinity Run ([1.2, 3.5]): $\sqrt{(1.2-3.0)^2+(3.5-2.7)^2}=1.97$
        \item Adidas UltraBoost ([2.0, 3.2]): $\sqrt{(2.0-3.0)^2+(3.2-2.7)^2}=1.12$
        \item Reebok Floatride ([3.1, 2.9]): $\sqrt{(3.1-3.0)^2+(2.9-2.7)^2}=0.22$
        \item ASICS Gel-Kayano ([2.5, 3.0]): $\sqrt{(2.5-3.0)^2+(3.0-2.7)^2}=0.58$
\end{enumerate}
Reebok Floatride is the closest match with a distance of 0.22, and it also meets the price criteria (under \$100).

\textbf{Euclidean Distance Visualization}

Visualize the embeddings on a 2D plane.

\begin{figure}[h!]
    \centering
    \includegraphics[width=1.10\linewidth]{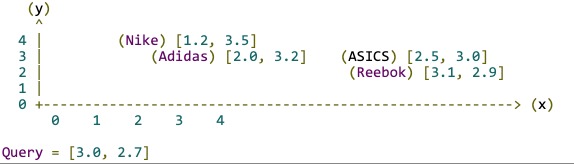}
    \caption{Distance Visualization}
    \label{fig:Distance Visualization}
\end{figure}

\textbf{The Result}
After applying the similarity search and filtering by price.

\begin{itemize}
    \item Top Recommendation: Reebok Floatride (Distance = 0.22, Price = \$90)
    \item Runner-Up: ASICS Gel-Kayano (Distance = 0.58, Price = \$110)
    \item Other products either fall short in semantic relevance or exceed the budget.
    \item The vector database delivers these results quickly, ensuring that even subtle nuances in user intent are matched with the most appropriate product.
\end{itemize}

\textbf{Key Takeaways}
\begin{enumerate}
    \item Semantic vs. Keyword: By converting text into embeddings, the system captures the true meaning behind queries and product descriptions, not just exact string matches.
    \item Vector Databases: Designed to store and query high-dimensional vectors, they provide efficient and scalable semantic search capabilities.
    \item Hybrid Queries: Combining semantic search with metadata filters (such as price) yields highly relevant and user-specific recommendations.
    \item Visualization and Intuition: Using Euclidean distance in a 2D plot helps illustrate how semantic similarity is determined, making the concept more accessible.
\end{enumerate}

Putting all these steps together, we start by recognizing the limitations of keyword-based searches, which often fail to capture the user’s true intent. We then prepare our data by converting product descriptions into vector embeddings using a language model, ensuring each item’s semantic attributes are preserved. These embeddings are stored in a vector database, specifically designed for efficient similarity searches on high-dimensional data. When a user submits a query, we generate an embedding for that query and perform a nearest neighbor lookup against our stored product vectors. Next, we refine the results by applying any necessary metadata filters such as price or brand and rank them based on their computed distance or similarity scores. Visualizing the data in a 2D plot (for demonstration purposes) helps illustrate how these distances translate into meaningful matches. Ultimately, the system delivers contextually relevant product recommendations that not only match the user’s stated preferences but also account for subtle nuances, creating a more intuitive and effective search experience.
\\

\textbf{Tip! Vector indexing for better performance}
To optimize similarity searches, leverage specialized indexing structures such as IVF, HNSW, or PQ-based approaches. These indexes enable the vector database to quickly narrow down candidate vectors, significantly reducing query times while maintaining high accuracy even at scale.
\\

\begin{itemize}
\item \textbf{Step 6: AI Model Execution \& Prompt Engineering}
\end{itemize}

In this step, the application constructs a carefully engineered prompt by integrating the user’s query, relevant context retrieved from prior interactions, and any auxiliary knowledge necessary to enhance the response. This comprehensive prompt is then passed to the large language model, which synthesizes the aggregated information to generate a contextually accurate and coherent answer. By leveraging both the user’s question and the retrieved context, the LLM is able to provide a tailored, informative response that directly addresses the user’s needs.

\textbf{Database Role:}

\begin{itemize}
    \item No direct database query happens at this step, but previously retrieved data ensures that the AI model generates a well-informed response.
    \item In some applications, embeddings generated at this stage are also stored in a vector database, streamlining future similarity searches and ensuring the LLM has immediate access to contextually relevant information for subsequent prompts.
\end{itemize}

\begin{itemize}
    \item \textbf{Step 7: Updating Conversation State \& Storing User Interactions}
\end{itemize}
    
After the large language model generates a response, the application must update its records to reflect the latest exchange. This involves capturing the user’s question, the LLM’s response, and any relevant metadata such as timestamps, user identifiers, or session IDs. By preserving these details, the system maintains a comprehensive conversation history, which can be used for future context retrieval, analytics, and model improvements.

\textbf{Database Role:}

\begin{itemize}
    \item A Key-Value or Document logs the user query, AI-generated response, and metadata (timestamps, feedback, etc.).
    \item In enterprise settings, structured responses may be logged in a relational database for future audits and analytics.
\end{itemize}

\begin{itemize}
    \item \textbf{Step 8: Response Generation \& Delivery}
\end{itemize}
    
Once the large language model (LLM) has produced its output, the system finalizes and delivers the response to the user. This stage can involve several sub-processes:
\begin{itemize}
    \item \textbf{Post-Processing and Formatting:}
    The raw LLM output may be refined or formatted according to the application’s requirements (e.g., ensuring consistency with UI guidelines or applying a content filter to remove sensitive data).
    Additional metadata, such as response confidence scores or relevant references, can be appended for user clarity or internal logging.
    \item \textbf{Delivery Mechanism:}
    The response is then sent through the appropriate communication channel such as a web interface, mobile application, or API endpoint to reach the user.
    In some cases, the system may adapt the presentation style based on the user’s device or preferences (e.g., text-only versus rich media).
    \item \textbf{Logging and Analytics:}
    Simultaneously, the system may log details of the interaction like response time, content, or user feedback in a database for performance monitoring and future analysis.
    These records enable developers to assess the LLM’s effectiveness, detect issues, and iteratively refine prompts or model parameters.
    By handling the final output with care, the application ensures users receive a coherent, context-aware answer while also capturing essential data for continuous improvement.
\end{itemize}

\textbf{Database Role:}
\begin{itemize}
    \item If the system includes caching mechanisms, frequently used responses may be fetched from a low-latency database to optimize performance.
    \item The system may also use historical user interactions to refine responses over time via continuous learning and fine-tuning.
\end{itemize}

By understanding the unique characteristics of conversational, situational, and semantic contexts in GenAI applications helps ensure the right database is used for each type of data, optimizing performance and scalability. By selecting the most appropriate database for each context, applications can achieve faster query responses, more relevant results, and improved overall efficiency in managing diverse data types. This approach brings value by aligning database capabilities with the specific needs of the context, resulting in more accurate, personalized, and context-aware user experiences.
Comparing these context types clarifies their distinct data requirements and retrieval patterns, guiding more precise database selection. By aligning each context with a fitting storage solution, teams can achieve better performance, scalability, and reliability. Ultimately, this approach maximizes the quality and speed of GenAI responses, ensuring they are both contextually rich and efficient.

Below is a comparative breakdown of conversational context, situational context, and semantic context in GenAI applications, highlighting the most suitable database types, the reasoning behind these choices, and relevant open-source examples.

\section*{Comparison of Database Contexts}

\begin{description}
    \item[Conversational Context] Stores chat history, user interactions, and messages exchanged with the AI. Maintains continuity for coherent responses across sessions. This context is dynamic, unstructured, and requires rapid read/write operations for real-time updates.
\end{description}
    
        \textbf{Database Type:} Redis-Key-Value, MongoDB-Document, PostgreSQL-Relational if strong consistency is required.
        
        \textbf{Reason:} \textbf{Flexibility \& Speed:} Schema-less storage supports varying conversation formats and enables fast updates and retrievals in real-time.
        
        \textbf{Transaction Pattern:} 
        
        \begin{enumerate}
            \item \textbf{High-Frequency Writes:} Continuous insertion of new messages as conversations evolve.
            \item \textbf{Frequent Reads:} Quick retrieval of recent or historical chat logs to maintain context.
            \item \textbf{High Concurrency:} Multiple simultaneous user sessions, each generating or reading messages in real time.
        \end{enumerate}

\begin{description}
    \item[Situational Context] Contains structured data like user profiles, operational metrics, and domain-specific information used to enrich and personalize responses.] 
\end{description}
    
        \textbf{Database Type:} Relational-MySQL, PostgreSQL, Delta Lakehouse-Apache Hudi
        
        \textbf{Reason:} \textbf{Structure \& Consistency:} Well-defined schemas ensure reliable storage, consistency, and robust querying of user and operational data.
        
        \textbf{Transaction Pattern:} 
        
        \begin{enumerate}
            \item \textbf{Frequent Reads:} Repeated lookups of user profiles, preferences, or operational data for personalization and decision-making.
            \item \textbf{Occasional Writes/Updates:} Periodic modifications to user information or metrics (e.g., profile changes, updated analytics).
            \item \textbf{Concurrent Access:} Various services and user sessions accessing and updating structured data concurrently.
        \end{enumerate}
    
    \begin{description}
        \item[Semantic Context] Manages high-dimensional embeddings and vectorized data for similarity searches, enabling semantic understanding beyond keyword matching.
    \end{description}
    
        \textbf{Database Type:} Vector Database-Milvus[25], FAISS(library, often embedded in custom solutions)[26], Vespa[27], pgvector extension[28]
        
        \textbf{Reason:} \textbf{Semantic Retrieval:} Optimized for storing and retrieving vector embeddings, enabling efficient similarity searches and context-aware responses.
        
        \textbf{Transaction Pattern:} 
        
        \begin{enumerate}
            \item \textbf{Frequent Reads:} Frequent similarity searches on stored embeddings to retrieve contextually relevant information.
            \item \textbf{Occasional Writes:} Inserting or updating embeddings when new data or content becomes available.
            \item \textbf{Optimized for Vector Operations:} Specialized indexing (e.g., IVF, HNSW) to handle high-dimensional queries efficiently, with generally lower concurrency than chat-based contexts.
        \end{enumerate}

\section*{Conclusion}
In conclusion, by leveraging the distinct power of each context conversational, situational, and semantic GenAI applications can deliver responses that are both relevant and intuitive. Each context type requires specialized database solutions, ensuring not only the accurate storage and retrieval of information but also the ability to process data at scale. As we’ve seen, integrating the right combination of relational, document, and vector databases for each use case enables AI systems to offer more refined and contextually aware interactions. This strategic approach enhances the overall user experience, making AI-driven applications smarter, faster, and more reliable for end-users. 
In conclusion, each type of context conversational, situational, and semantic plays a unique role in delivering comprehensive, accurate responses within GenAI applications. By leveraging the appropriate database solutions for each context, AI systems can combine real-time conversation logs, external domain data, and vectorized semantic knowledge into a cohesive and responsive experience. As we turn our attention to semantic context, we’ll see how vector databases and similarity searches bring deeper meaning to user queries, further enhancing the quality and relevance of the AI’s outputs.

\textbf{Footnote}: This work is independent and not affiliated with my role at Amazon.

\bibliographystyle{IEEEtran} 
\bibliography{references,mybib} 

\cite{gpt4, gemini, instruction-follow, rag, topol_ai, ai_finance, clip, codex, faiss, pq, scale, model_versioning, clipper, one_size, vector_search, redis, mongodb, postgresql, mysql, hudi, milvus, faiss_lib, vespa, pgvector}

\end{document}